\documentclass[a4paper,11pt]{article}
\pdfoutput=1 

\usepackage{jheppub} 

\usepackage[T1]{fontenc} 
\def\arXiv#1{\href{http://arxiv.org/abs/#1}{arXiv:#1}}
\def\arXiv#1#2{\href{http://arxiv.org/abs/#1}{arXiv:#1}}
\def\arXivid#1#2{\href{http://arxiv.org/abs/#1/#2}{#1/#2}}
\def\doi#1#2{\href{http://doi.org/#1}{#2}}

\title{\boldmath Asymptotically AdS spacetimes with a timelike Kasner singularity}

\author{Jie Ren}

\affiliation{Racah Institute of Physics, The Hebrew University of Jerusalem, 91904, Israel}

\emailAdd{jie.ren@mail.huji.ac.il}

\abstract{Exact solutions to Einstein's equations for holographic models are presented and studied. The IR geometry has a timelike cousin of the Kasner singularity, which is the less generic case of the BKL (Belinski-Khalatnikov-Lifshitz) singularity, and the UV is asymptotically AdS. This solution describes a holographic RG flow between them. The solution's appearance is an interpolation between the planar AdS black hole and the AdS soliton. The causality constraint is always satisfied. The entanglement entropy and Wilson loops are discussed. The boundary condition for the current-current correlation function and the Laplacian in the IR is examined. There is no infalling wave in the IR, but instead, there is a normalizable solution in the IR. In a special case, a hyperscaling-violating geometry is obtained after a dimensional reduction.}

\begin{document} 
\maketitle
\flushbottom

\section{Introduction}
\label{sec:intro}
The planar AdS black hole plays a basic role in the AdS/CFT correspondence, where a strongly coupled quantum field theory in the AdS boundary is mapped to a classical gravity in the bulk. Another basic solution is the AdS soliton, which appears as a double Wick rotation of the AdS black hole. At a critical temperature, there is a confinement-deconfinement phase transition between the AdS soliton and the AdS black hole \cite{Witten:1998zw}. The AdS$_4$ black hole in Poincar\'{e} coordinates is
\begin{equation}
ds^2=\frac{L^2}{r^2}\left(-fdt^2+\frac{dr^2}{f}+dx^2+dy^2\right),
\end{equation}
where $f=1-(r/r_0)^3$. The AdS boundary is at $r=0$, and the horizon is at $r=r_0$.

The following ``deformation'' of the planar AdS black hole is {\it still an exact solution to Einstein's equations}:
\begin{equation}
ds^2=\frac{L^2}{r^2}\left(-f^\alpha dt^2+\frac{dr^2}{f}+f^\beta dx^2+f^\gamma dy^2\right),\label{eq:sol}
\end{equation}
where $\alpha$, $\beta$, and $\gamma$ are constants satisfying the Kasner conditions
\begin{equation}
\alpha+\beta+\gamma = 1,\qquad
\alpha^2+\beta^2+\gamma^2 = 1.
\end{equation}
This solution is obtained by mapping the time evolution of a Bianchi type I Universe to a holographic RG flow, as explained in appendix~\ref{sec:other}. At first sight, this solution interpolates between the AdS black hole and the AdS soliton:
\begin{itemize}
\item $\alpha=1$, $\beta=\gamma=0$. In this case, $r=r_0$ is the (finite temperature) horizon of the AdS black hole.
\item $\alpha=\beta=0$, $\gamma=1$. In this case, $r=r_0$ is regular, and the solution is the AdS soliton.
\end{itemize}

However, when $\alpha\neq 1$, the IR geometry at $r=r_0$ is completely different from a finite temperature horizon. By defining $\bar{r}=\sqrt{r_0-r}$, the IR geometry is
\begin{equation}
ds^2=-\bar{r}^{2\alpha}dt^2+d\bar{r}^2+\bar{r}^{2\beta}dx^2+\bar{r}^{2\gamma}dy^2.
\end{equation}
This is the Kasner solution \cite{Kasner:1921} in Ricci-flat spacetime after a double Wick rotation. At $\bar{r}=0$ ($r=r_0$), there is a naked singularity, and we call it {\it timelike Kasner singularity}, as distinguished to the spacelike Kasner singularity in cosmology. The Kasner singularity is the less generic case of the BKL (Belinski-Khalatnikov-Lifshitz) singularity, while the generic BKL singularity has oscillatory behaviors \cite{Lifshitz:1963,Belinsky:1970}. The timelike BKL singularity has been recently studied in \cite{Shaghoulian:2016umj} as an IR fixed in AdS/CFT; see \cite{Khalatnikov:1978,Parnovsky:1980} for some related earlier works. The solution \eqref{eq:sol} is asymptotically AdS, and thus describes an RG flow from the AdS boundary as the UV, to the timelike Kasner singularity as the IR.

Naked singularities are not uncommon in AdS/CFT. We have plenty of examples of supergravity systems with naked singularities from consistent truncations of the 10D superstring or 11D M-theory \cite{Cvetic:1999xp}. The IR geometries are hyperscaling-violating geometries.  The parameters of the hyperscaling-violating geometries are constrained by the Gubser criterion \cite{Gubser:2000nd}. When the Gubser criterion is satisfied, the geometry can be obtained as an extremal limit of a finite temperature black hole. The extremal limit is at either $T\to 0$ or $T\to\infty$; a property of the latter case is that the spectral functions in these geometries can have a hard gap \cite{Charmousis:2010zz,Kiritsis:2015oxa}.

According to \cite{Iizuka:2012wt} (and also \cite{Glorioso:2015vrc}), anisotropic geometries with a regular horizon do not exist in pure gravity, which implies that the BKL singularity does not satisfy the Gubser criterion without matter fields. However, the Gubser criteron is a sufficient but not necessary condition to justify a naked singularity \cite{Gubser:2000nd}. Justifications for the timelike BKL singularity have been studied in \cite{Shaghoulian:2016umj}, including the causality constraint \cite{Kleban:2001nh,Bak:2004yf,Gao:2000ga} and the accessibility to the IR \cite{Engelhardt:2013tra}. We will show that these two conditions are always satisfied for the anisotropic solution. Beside, we will examine another constraint related to the boundary conditions at the IR. All these suggest that the timelike BKL singularity is relatively well behaved.

\begin{figure}
  \centering
  \includegraphics{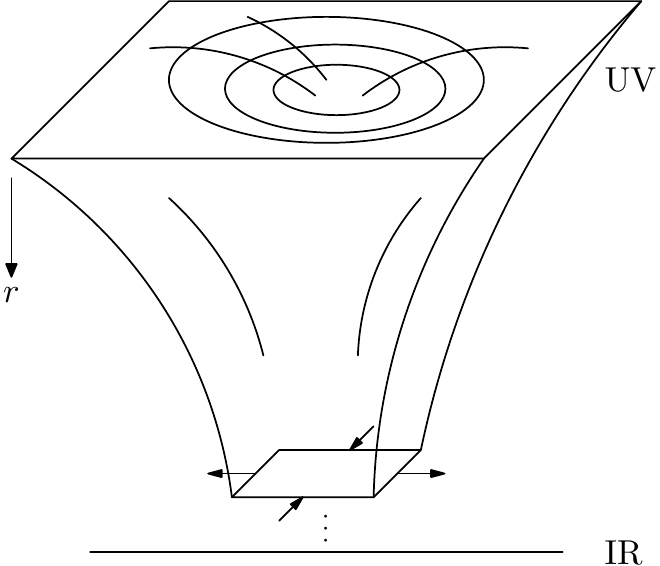}
  \caption{\label{fig:geometry} Schematic plot of the geometry, i.e., the RG flow from UV to IR. The UV is asymptotically AdS. As the IR is approached, some directions will be contracting, while some directions will be expanding. The Kasner singularity in the IR is line-like or hyperplane-like.}
\end{figure}

To my taste, the remarkable simplicity of the metric \eqref{eq:sol} alone demands that it must be interesting physically. We will explore various holographic aspects of the solution \eqref{eq:sol} and its higher dimensional generalizations, including one-point and two-point functions, the entanglement entropy, and Wilson loops. In the bulk, the geometry is anisotropic, as illustrated in figure~\ref{fig:geometry}. From the phenomenological point of view, there are anisotropic materials in condensed matter physics, for example, \cite{Ito:1991}. A key physical quantity to calculate is the conductivity. For a black hole, we impose the infalling wave boundary condition in the IR for perturbation equations. For the timelike Kasner singularity, however, the correct boundary condition in the IR is normalizability.

We can obtained a hyperscaling-violating geometry after a Kaluza-Klein (KK) reduction of a special case of the anisotropic solution. The lower-dimensional hyperscaling-violating geometry cannot be connected to a finite temperature black hole; the Gubser criterion is marginally violated, while the null energy condition is marginally satisfied.

This paper is organized as follows. In the next section, we present the anisotropic solution in arbitrary dimensions, obtain the stress tensor of the boundary CFT and examine the energy conditions. In section~\ref{sec:probe}, we show that the causality constraint is always satisfied by analyzing the null geodesics. We also examine the entanglement entropy and Wilson loops. In section~\ref{sec:conductivity}, we examine the boundary conditions in the IR for solving correlation functions. In section~\ref{sec:hyperscaling}, we obtain a hyperscaling geometry after a KK reduction for a special case of the anisotropic solution. Finally, we summarize the properties of this solution and discuss some further issues.

\section{Anisotropic solution and boundary stress tensor}
\label{sec:solution}
We consider the AdS$_{n+2}$ action
\begin{equation}
S=\int d^{n+2}x\sqrt{-g}\left(R+\frac{n(n+1)}{L^2}\right),\label{eq:action}
\end{equation}
where $L$ is the AdS radius. The Einstein's equations have the following solution:
\begin{equation}
ds^2=\frac{L^2}{r^2}\left(-f^{p_t}dt^2+\frac{dr^2}{f}+\sum_{i=1}^n f^{p_i}dx_i^2\right),
\label{eq:soln}
\end{equation}
where
\begin{equation}
f=1-\left(\frac{r}{r_0}\right)^{n+1},
\end{equation}
and $p_i$ ($p_0\equiv p_t$) satisfies the Kasner conditions
\begin{equation}
\sum_{i=0}^n p_i=1,\qquad \sum_{i=0}^n p_i^2=1.\label{eq:kasner}
\end{equation}
The IR limit of the geometry is at $r=r_0$. We will set $L=1$ and $r_0=1$ throughout this paper. Obviously, the Kasner conditions require $|p_i|\leq 1$.

By defining $\bar{r}=\sqrt{r_0-r}$, the IR geometry of \eqref{eq:soln} is
\begin{equation}
ds^2=-\bar{r}^{2p_t}dt^2+d\bar{r}^2+\sum_{i=1}^n\bar{r}^{2p_i}dx_i^2.\label{eq:ksoln}
\end{equation}
This is the Kasner solution in Ricci-flat spacetime after a double Wick rotation, which implies that the cosmological constant is irrelevant in the IR. There is a timelike Kasner singularity at $\bar{r}=0$.\footnote{The metric \eqref{eq:ksoln} describes the less generic case of the timelile BKL singularity. In the generic case, the (3+1)-dimensional ansatz is
\begin{equation}
ds^2=d\bar{r}^2+(-a^2l_\mu l_\nu+b^2m_\mu m_\nu+c^2n_\mu n_\nu)dx^\mu dx^\nu.
\end{equation}
Please refer to \cite{Shaghoulian:2016umj,Parnovsky:1980} for studies about this case; the geometry has oscillatory behaviors by swapping exponents of \eqref{eq:ksoln} toward the singularity.}
When the IR is approached, the $p_i>0$ directions will be contracting, while the $p_i<0$ directions will be expanding. We may also have $p_i=0$ directions.

We can obtain the stress-energy tensor of the boundary CFT by the holographic renormalization \cite{deHaro:2000vlm,Skenderis:2002wp}. We will briefly review the procedure, and apply it to the anisotropic solution. Write the metric in Fefferman-Graham coordinates
\begin{equation}
ds^2=\frac{L^2}{z^2}\left(dz^2+g_{\mu\nu}(x^\mu,z)dx^\mu dx^\nu\right),\label{eq:GF}
\end{equation}
where the near boundary $z\to 0$ behavior of $g_{\mu\nu}(x^\mu,z)$ is
\begin{equation}
g(x^\mu,z)=g_{(0)}+z^2g_{(2)}+\cdots+z^{n+1}g_{(n+1)}+h_{(n+1)}z^{n+1}\log z^2+\mathcal{O}(z^{n+2}).
\end{equation}
The stress-energy tensor is
\begin{equation}
\langle T_{\mu\nu}\rangle=\frac{(n+1)L^n}{16\pi G_N}g_{(n+1)\mu\nu}+X_{\mu\nu}[g_{(k)}],
\label{eq:Tmunu}
\end{equation}
where $X_{\mu\nu}[g_{(k)}]$ is a function of $g_{(k)}$ with $k<n+1$, and it reflects the conformal anomalies of the boundary CFT \cite{deHaro:2000vlm}. In our case, we will see that $X_{\mu\nu}=0$.

By comparison of the metric \eqref{eq:soln} and \eqref{eq:GF}, the relation between $r$ and $z$ is ($r_0=1$)
\begin{equation}
r=z\left(\frac{2}{1+z^{n+1}}\right)^\frac{2}{n+1}.
\end{equation}
Thus, we obtain the geometry \eqref{eq:soln} in Fefferman-Graham coordinates:
\begin{equation}
ds^2=\frac{L^2}{z^2}\left(dz^2-f^{2p_t}h\,dt^2+\sum_{i=1}^nf^{2p_i}h\,dx_i^2\right),
\end{equation}
where
\begin{equation}
f=\frac{1-(z/z_0)^{n+1}}{1+(z/z_0)^{n+1}},\qquad h=\left[1+(z/z_0)^{n+1}\right]^\frac{4}{n+1}.
\end{equation}
The boundary stress-energy tensor can be straightforwardly obtained by \eqref{eq:Tmunu} (no summation for $T_{ii}$):
\begin{align}
&\bar{\epsilon}\equiv\langle T_{00}\rangle=\frac{L^n}{4\pi G_N}\frac{(n+1)p_t-1}{z_0^{n+1}},\\
&\bar{p}_i\equiv\langle T_{ii}\rangle=\frac{L^n}{4\pi G_N}\frac{1-(n+1)p_i}{z_0^{n+1}},
\end{align}
where $\bar{\epsilon}$ is the energy density, and $\bar{p}_i$ is the pressure along the $i$ direction ($1\leq i\leq n$). By the first Kasner condition in \eqref{eq:kasner}, the stress-energy tensor is traceless.

The energy conditions for the boundary CFT give nontrivial constraints for the parameters $p_t$ and $p_i$. First, the energy density $\bar{\epsilon}\geq 0$ gives\,\footnote{Note that these energy conditions are for classical field theories. For a quantum field theory, the energy density can be negative due to the Casimir effect. For example, the energy density of the CFT dual to the AdS soliton is negative. Thus, the constraints discussed below are not strictly necessary.}
\begin{equation}
p_t\geq\frac{1}{n+1}.\label{eq:posit}
\end{equation}
The null energy condition ($\bar{\epsilon}+\bar{p}_i\geq 0$ for all $i$) gives
\begin{equation}
p_t\geq p_i\quad\text{for all $i$}.\label{eq:NEC}
\end{equation}
The first Kasner condition in \eqref{eq:kasner} combined with \eqref{eq:NEC} implies \eqref{eq:posit}. Furthermore, when $p_t\geq p_i$ is satisfied for all $i$, the weak and the strong energy conditions are also satisfied, but the dominant energy condition ($\bar{\epsilon}\geq |\bar{p}_i|$ for all $i$) needs an extra condition $p_t+p_i\geq 2/(n+1)$ for all $i$.

Take the AdS$_4$ solution \eqref{eq:sol} as an example, and assume $\beta\geq\gamma$ without loss of generality. The null, weak or strong energy condition combined with the Kasner conditions implies
\begin{equation}
2/3\leq\alpha\leq 1,\qquad 0\leq\beta\leq 2/3,\qquad -1/3\leq\gamma\leq 0.
\end{equation}
The dominant energy condition is more stringent, and implies
\begin{equation}
(1+\sqrt{3})/3\leq\alpha\leq 1,\qquad 0\leq\beta\leq 1/3,\qquad (1-\sqrt{3})/3\leq\gamma\leq 0.
\end{equation}

The null energy condition ensures $g_{tt}\to 0$, and that the product of all $g_{ii}$ vanishes at the IR, although there are contracting and expanding spatial directions.

\section{Geodesics and extremal surfaces}
\label{sec:probe}
\subsection{Geodesics}
Geodesics and extremal surfaces are bulk probes in AdS/CFT \cite{Hubeny:2012ry}. We will study geodesics first. The null geodesics are closely related to the causality constraint. Let $\lambda$ be the affine parameter along the geodesics, and the geodesics are extrema of the action
\begin{equation}
S'=\int d\lambda\, g_{\mu\nu}\dot{x}^\mu\dot{x}^\nu=\int d\lambda\,\frac{1}{r^2}\left(-f^{p_t}\dot{t}^2+\frac{\dot{r}^2}{f}+\sum_{i=1}^nf^{p_i}\dot{x}_i^2\right),
\end{equation}
where the dots are with respect to $\lambda$. Form this action, we can see that the energy $E$ and momenta $P_i$ are conserved quantities (no summation)
\begin{equation}
E=\frac{f^{p_t}}{r^2}\dot{t},\qquad P_i=\frac{f^{p_i}}{r^2}\dot{x}_i.\label{eq:conserv}
\end{equation}
The geodesic equation for $r(\lambda)$ can be obtained by $g_{\mu\nu}\dot{x}^\mu\dot{x}^\nu=\kappa$, where $\kappa=+1$, $0$, $-1$ for spacelike, null, and timelike geodesics, respectively. The equation for $r$ is
\begin{equation}
\dot{r}^2+V_\text{eff}(r)=0,\qquad V_\text{eff}(r)=r^2f\left(-\kappa-\frac{r^2}{f^{p_t}}E^2+\sum_{i=1}^n\frac{r^2}{f^{p_i}}P_i^2\right).
\end{equation}
Since $\dot{r}^2\geq 0$, it requires $V_\text{eff}\leq 0$. The deepest point in the bulk (largest $r$) that a geodesic can reach is given by the smallest root of $V_\text{eff}(r)=0$. In the IR limit, there are three cases:
\begin{itemize}
\item The AdS black hole ($p_t=1$). $V_\text{eff}\to -E^2$, which implies that the horizon is attractive.
\item The AdS soliton (one of $p_i=1$). $V_\text{eff}\to P_i^2$, which implies that its tip is repulsive.
\item The Kasner singularity ($|p_t|<1$ and all $|p_i|<1$). $V_\text{eff}\to 0$, which implies that the singularity is neither attractive nor repulsive.
\end{itemize}

Assume a photon is propagating along the $i$ direction. Since $\kappa=0$ and $E=P_i$ for null geodesics, we have $V_\text{eff}\leq 0$ if $p_t>p_i$; $V_\text{eff}\geq 0$ if $p_t<p_i$; and $V_\text{eff}=0$ if $p_t=p_i$. Therefore, the null geodesics can get access to the bulk if $p_t>p_i$. The spacelike geodesics can reach deeper in the bulk than null geodesics \cite{Hubeny:2012ry}, and do not require this constraint.

We will use the causality constraint \cite{Kleban:2001nh} to justify the solution \eqref{eq:soln}, as suggested by \cite{Shaghoulian:2016umj}. For two spatial points on the boundary, the causality constraint requires that the fastest null geodesic connecting these two points always lies in the boundary, namely
\begin{equation}
\int_\text{bulk}dt\geq\int_\text{bry}dt.
\end{equation}
If this is satisfied, spacelike separated points along the boundary are also spacelike separated throughout the bulk. Assume a photon is propagating along a null geodesic with two ends on the AdS boundary in the $i$ direction, and consider a small interval of the affine parameter $\Delta\lambda$. From \eqref{eq:conserv}, we can see that
\begin{equation}
\frac{\Delta t}{\Delta x}=\frac{E/f^{p_t}}{P_i/f^{p_i}}=\frac{P_i}{E}\cdot\frac{E^2/f^{p_t}}{P_i^2/f^{p_i}}\geq 1,
\end{equation}
where we used $V_\text{eff}\leq 0$, which gives $E^2/f^{p_t}\geq P_i^2/f^{p_i}$. In the above inequality, $\Delta t$ is the time along the null geodesic in the bulk, and $\Delta x$ equals the time along a null geodesic in the boundary. By integrating along $\lambda$, we can demonstrate that the causality constraint is always satisfied.

In \cite{Shaghoulian:2016umj}, it was argued that the causality constraint is satisfied as long as $p_t>p_i$, by analyzing the IR geometry \eqref{eq:ksoln}. With the asymptotically AdS solution as a full RG flow, we conclude that $p_t>p_i$ is the condition that the null geodesics can probe into the bulk, while the causality constraint does not require $p_t>p_i$, and is always satisfied.

\subsection{Entanglement entropy and Wilson loops}
\label{sec:Wilson}
According to \cite{Shaghoulian:2016umj,Engelhardt:2013tra}, another condition for a physical geometry is that extremal surfaces anchored at the AdS boundary should not have any barrier to access the IR, which requires the extrinsic curvature to be positive. For the radial normal vector $\hat{n}=r\sqrt{f}\,\partial_r$, the extrinsic curvature is
\begin{equation}
K=h^{\mu\nu}\nabla_\mu \hat{n}_\nu=\frac{rf'-2(n+1)f}{2\sqrt{f}},
\end{equation}
which is always positive for all $0\leq r\leq r_0$ and is independent of the anisotropy.

{\it Entanglement entropy.} The entanglement entropy is calculated from codimension-2 extremal surfaces in the bulk by the Ryu-Takayanagi formula \cite{Ryu:2006bv,Ryu:2006ef}. Consider a strip with width $l$ in the direction $x\equiv x_1$, and infinite width in other directions. The metric of the codimension-2 surface in the bulk can be written as
\begin{equation}
ds^2=\frac{1}{r^2}\left((f^{-1}+f^\beta x'(r)^2)dr^2+\sum_{i=2}^{n}f^{p_i}dx_i^2\right).
\end{equation}
The area of the bulk surface is
\begin{equation}
S=\int d^nx\sqrt{-h}=A_{n-1}\int\frac{1}{r^n}\sqrt{(f^{-1}+f^\beta x'^2)f^{1-\alpha-\beta}}\,dr,
\label{eq:action2}
\end{equation}
where $A_{n-1}$ is the area of the other $n-1$ dimensions. In the $n=2$ (AdS$_4$) case, this is equivalent to a holographic Wilson loop whose worldsheet coordinates are $\sigma^1=r$ and $\sigma^2=y$.

Let $r_m$ be the maximal value of $r$ for a connected surface. At $r=r_m$, we have $x'(r)\to\infty$. Since the Lagrangian does not contain $x(r)$, the following quantity is conserved:
\begin{equation}
\frac{f^{(1-\alpha)/2}}{r^n}\frac{x'}{\sqrt{x'^2+f^{-1-\beta}}}=\frac{f_m^{(1-\alpha)/2}}{r_m^n},
\label{eq:conserv2}
\end{equation}
where $f_m=f(r_m)$. Solving $x'$ from \eqref{eq:conserv2}, and plugging it into \eqref{eq:action2}, we obtain
\begin{equation}
S=A_{n-1}\int_\epsilon^{r_m}\frac{f^{-(\alpha+\beta)/2}}{r^n}\sqrt{\frac{(f/f_m)^{1-\alpha}}{(f/f_m)^{1-\alpha}-(r/r_m)^4}}\,dr,
\end{equation}
which has been regulated by a UV cutoff at $r=\epsilon$. The value of $r_m$ is determined by integrating $x'$ solved from \eqref{eq:conserv2}:
\begin{equation}
\frac{l}{2}=\int_0^{r_m}\frac{(r/r_m)^{n/2}f^{-(1+\beta)/2}}{\sqrt{(f/f_m)^{1-\alpha}-(r/r_m)^4}}\,dr.
\end{equation}

By defining $u=r/r_m$, we can write the above integral in terms of \eqref{eq:lrm} in appendix~\ref{sec:int}. For the anisotropic solution, we have $1-\alpha>0$, so there is a maximal length $l_\text{max}$, similar to \cite{Klebanov:2007ws}. When $l<l_\text{max}$, the connected surface exists. When $l>l_\text{max}$, the connected surface does not exist. There may be a critical length $l_\text{crit}$ \cite{Klebanov:2007ws}. When $l_\text{crit}<l<l_\text{max}$, the disconnected surface will dominate:
\begin{equation}
S^\text{(disconn)}=A_{n-1}\int_\epsilon^1\frac{f^{-(\alpha+\beta)/2}}{r^n}.
\end{equation}

We can write an effective potential for a zero-energy particle from \eqref{eq:conserv2}:
\begin{equation}
\left(\frac{dr}{dx}\right)^2+V_\text{eff}(r)=0,\qquad V_\text{eff}(r)=f^{1+\beta}\left(1-\frac{(f/f_m)^{1-\alpha}}{(r/r_m)^{2n}}\right)
\end{equation}
For the anisotropic solution, we have $1-\alpha>0$. Therefore, there is always a maximal length $l_\text{max}$ for the connected surface, according to the analysis of \eqref{eq:Veff1} below.

\begin{figure}
  \centering
  \includegraphics{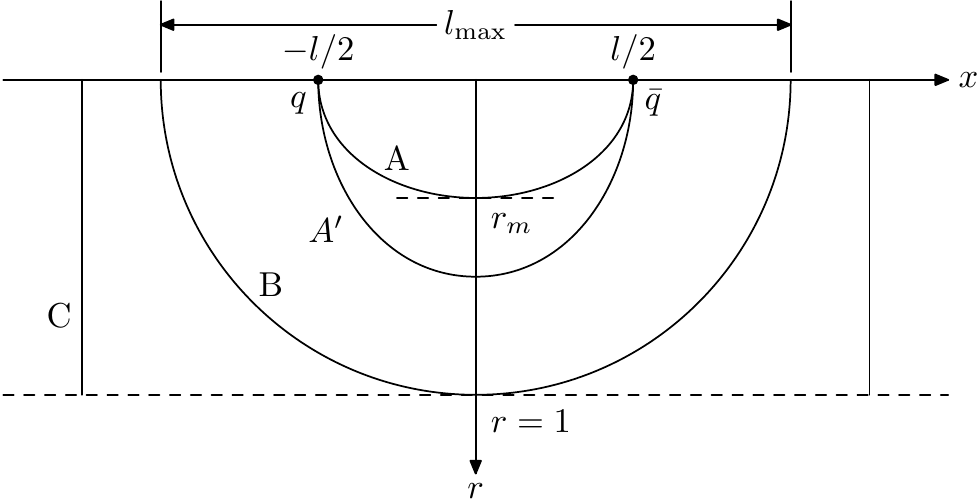}
  \caption{\label{fig:Wilson} Schematic plot of a string connecting a quark and an antiquark on the AdS boundary. When $\alpha+\beta>0$, there are two connected solutions $A$ and $A'$ for a given quark-antiquark separation $l<l_\text{max}$. There is also a disconnected solution $C$. This plot can also be thought of as a qualitative representation of extremal surfaces.}
\end{figure}
\begin{figure}
  \centering
  \includegraphics{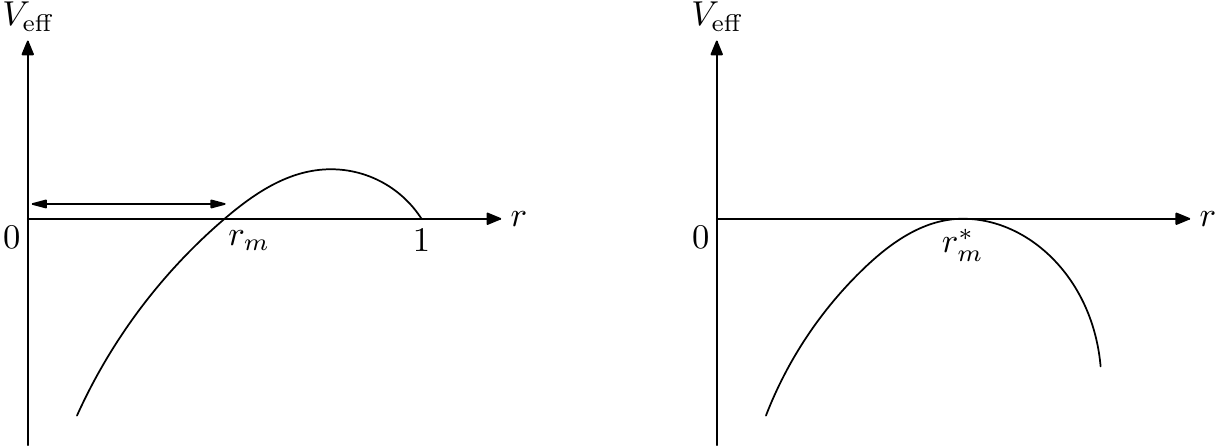}
  \caption{\label{fig:potential} Schematic plot of the effective potential in \eqref{eq:Veff1}. The left plot is for the case $\alpha+\beta>0$, in which a zero-energy particle will turn around at $r=r_m$ at a finite ``time'' ($l_\text{max}$). The right plot is for the case $\alpha+\beta<0$ and $r_m=r_m^*$, in which the zero-energy particle spends an infinitely long ``time'', when it approaches $r=r_m^*$.}
\end{figure}

{\it Wilson loops.} We want to calculate the quark-antiquark potential to see whether the anisotropic solution describes a confined or deconfined state. The quark-antiquark potential can be calculated by a holographic Wilson loop \cite{Maldacena:1998im}. Consider a string connecting a quark and an antiquark on the AdS boundary in the $x\equiv x_1$ direction, separated by $l$. We choose the worldsheet coordinate as $\sigma^0=t$ and $\sigma^1=r$, and the profile of the string is given by the function $x(r)$, as illustrated in figure~\ref{fig:Wilson}. Let $\alpha\equiv p_t$ and $\beta\equiv p_1$. The metric for the string worldsheet is\,\footnote{If the string ends in an arbitrary direction, the metric for the string worldwheet is
$$ds^2=\frac{1}{r^2}\left(-f^{p_t}dt^2+\Big(f^{-1}+\sum_{i=1}^nf^{p_i}x_i'(r)^2\Big)dr^2\right).$$}
\begin{equation}
ds^2=\frac{1}{r^2}\left(-f^\alpha dt^2+(f^{-1}+f^\beta x'(r)^2)dr^2\right).
\end{equation}
The Nambu-Goto action for the string is
\begin{equation}
S_\text{NG}=\int d^2\sigma\sqrt{-h}=T\int\frac{1}{r^2}\sqrt{f^\alpha(f^{-1}+f^\beta x'^2)}\,dr,
\label{eq:action1}
\end{equation}
where $T$ is the time duration of the Wilson loop.

Let $r_m$ be the maximal value of $r$ for a connected string. At $r=r_m$, we have $x'(r)\to\infty$. Since the Lagrangian does not contain $x(r)$, the following quantity is conserved:
\begin{equation}
\frac{f^{(\alpha+\beta)/2}}{r^2}\frac{x'}{\sqrt{x'^2+f^{-1-\beta}}}=\frac{f_m^{(\alpha+\beta)/2}}{r_m^2},
\label{eq:conserv1}
\end{equation}
where $f_m=f(r_m)$. Solving $x'$ from \eqref{eq:conserv1}, and plugging it into \eqref{eq:action1}, we obtain
\begin{equation}
S_\text{NG}=T\int_\epsilon^{r_m}\frac{f^{(\alpha-1)/2}}{r^2}\sqrt{\frac{(f/f_m)^{\alpha+\beta}}{(f/f_m)^{\alpha+\beta}-(r/r_m)^4}}\,dr,
\end{equation}
which has been regulated by a UV cutoff at $r=\epsilon$. The value of $r_m$ is determined by integrating $x'$ solved from \eqref{eq:conserv1}:
\begin{equation}
\frac{l}{2}=\int_0^{r_m}\frac{(r/r_m)^2f^{-(1+\beta)/2}}{\sqrt{(f/f_m)^{\alpha+\beta}-(r/r_m)^4}}\,dr.
\end{equation}
The quark-antiquark potential is given by $V_{q\bar{q}}(l)=S_\text{NG}/T$.

By defining $u=r/r_m$, we can write the above integral in terms of \eqref{eq:lrm} in appendix~\ref{sec:int}. When $\alpha+\beta>0$, $l$ have a maximum, $l_\text{max}$. When $l>l_\text{max}$, there is no connected string solution. This is similar to the Debye screening in the black hole geometry \cite{Rey:1998bq,Brandhuber:1998bs,Bak:2007fk}. For AdS$_4$, $\alpha+\beta=1-\gamma$, which is always positive as long as the geometry is not the AdS soliton. Therefore, the CFT dual to the AdS$_4$ anisotropic geometry is in a non-confining phase.

For $\alpha+\beta<0$, there will be a maximal value of $r_m$ denoted by $r_m^*$. We have $l(r_m^*)\to\infty$, which implies that we always have a connected string solution. In this case, the CFT dual to the anisotropic geometry is in a confined phase. This is possible in higher dimensions, i.e., $p_t+p_i<0$ for some $i$, if no other constraints are imposed.

There may be a critical length $l_\text{crit}$. When $l_\text{crit}<l<l_\text{max}$, the disconnected string contribution dominates \cite{Rey:1998bq,Brandhuber:1998bs,Bak:2007fk}, although the connected string solution exists.

A qualitative way to analyze the behavior of the string is by an effective potential, in which a zero-energy particle is moving, and the coordinate $x$ is the ``time'' \cite{Bak:2004yf}. Equation~\eqref{eq:conserv1} can be rewritten as
\begin{equation}
\left(\frac{dr}{dx}\right)^2+V_\text{eff}(r)=0,\qquad V_\text{eff}(r)=f^{1+\beta}\left(1-\frac{(f/f_m)^{\alpha+\beta}}{(r/r_m)^4}\right).
\label{eq:Veff1}
\end{equation}
\begin{itemize}
\item $\alpha+\beta>0$. For any $0<r_m<1$, we always have $V_\text{eff}\leq 0$ when $0\leq r\leq r_m$, as illustrated in the left plot of figure~\ref{fig:potential}. The zero-energy particle will turn around at $r=r_m$ at a finite ``time''. Equivalently, there is a maximal quark-antiquark separation $l_\text{max}$ for the connected string.
\item $\alpha+\beta<0$. There is a maximal value of $r_m$, such that for any $0<r_m<r_m^*$, we have $V_\text{eff}\leq 0$ when $0\leq r\leq r_m$. When $r_m=r_m^*$, the zero-energy particle spends an infinitely long ``time'', when it approaches the turning point, as illustrated in the right plot of figure~\ref{fig:potential}. The maximal value of $r_m$ is determined by $V_\text{eff}'(r_m^*)=0$:
\begin{equation}
r_m^*=\left(\frac{4}{4-(n+1)(a+b)}\right)^\frac{1}{n+1}.
\end{equation}
\end{itemize}

\section{Two-point correlation functions}
\label{sec:conductivity}
We will investigate the boundary conditions in the IR for solving two-point correlation functions. We begin with the current-current correlation function, which is responsible for the conductivity. See \cite{Hartnoll:2009sz,Hartnoll:2008kx} for applications of AdS/CFT to condensed matter physics. We perturb the system by an electric field in the $x\equiv x_1$ direction. For simplicity, we denote $\alpha\equiv p_t$ and $\beta\equiv p_1$. We assume $\alpha\geq\beta$ by the null energy condition of the boundary CFT.

{\it Conductivity.} To calculate the conductivity, we turn on a perturbation $\delta A_t=e^{-i\omega t}a_x$ around the background \eqref{eq:soln}, and the equation for $a_x$ is\,\footnote{The action is \eqref{eq:action} with a Maxwell term $-\frac{1}{4}F^2$.}
\begin{equation}
a_x''+\left((1-\beta)\frac{f'}{f}-\frac{n-2}{r}\right)a_x'+\frac{\omega^2}{f^{1+\alpha}}a_x=0.
\end{equation}
This equation is solved with appropriate boundary conditions in the IR to be discussed below. The asymptotic behavior of $a_x$ near the AdS boundary $r=0$ is
\begin{equation}
a_x=a_x^{(0)}+\cdots+a_x^{(n-1)}r^{n-1}+\cdots,
\end{equation}
where $a_x^{(0)}$ is the source, $a_x^{(n-1)}$ is the expectation value, and there are logarithmic terms when $n$ is odd. The DC conductivity is
\begin{equation}
\sigma(\omega)=\frac{a_x^{(n-1)}}{(n-1)i\omega a_x^{(0)}}.
\label{eq:sigma}
\end{equation}

After a change of variables by
\begin{equation}
\frac{d\xi}{dr}=\frac{1}{f^{(1+\alpha)/2}},\qquad \tilde{a}_x=r^{-\frac{n-2}{2}}f^{-\frac{\alpha+2\beta-1}{4}}a_x,
\end{equation}
we can obtain a Schr\"{o}dinger equation\,\footnote{The way to obtain a Schr\"{o}dinger equation can be found in appendix~C of \cite{Herzog:2007kh}, for example.}
\begin{equation}
-\frac{d^2\tilde{a}_x}{d\xi^2}+\tilde{V}(\xi)\tilde{a}_x=\omega^2\tilde{a}_x.
\label{eq:schr}
\end{equation}
The Schr\"{o}dinger potential is
\begin{equation}
\tilde{V}=\frac{(n-2)f^\alpha[nf-2(1-\beta)rf']}{4r^2}-\frac{1}{16}(\alpha+2\beta-1)f^{\alpha-1}[(\alpha-2\beta-1)f'^2+4ff''],
\end{equation}
where the derivatives are taken with respect to the original coordinate $r$. If $\tilde{V}\to\infty$ in both the UV and the IR, then the Schr\"{o}dinger equation has a discrete spectrum (and a hard gap), so does the conductivity. If $\tilde{V}\leq 0$ in either the UV or the IR, then the Schr\"{o}dinger equation has a continuous spectrum, and the conductivity is gapless.

In the IR limit, the Schr\"{o}dinger coordinate is
\begin{equation}
\xi=-\frac{2}{1-\alpha}(n+1)^{-\frac{1+\alpha}{2}}(1-r)^\frac{1-\alpha}{2}.\label{eq:xi}
\end{equation}
Since $\alpha<1$, the IR limit $r\to 1$ is always at $\xi=0$. The leading order of the Schr\"{o}dinger potential is
\begin{equation}
\tilde{V}=\frac{\nu^2-1/4}{\xi^2},\qquad \nu=\frac{\beta}{1-\alpha},
\end{equation}
which is independent of the dimension $n$. The general solution is a linear combination of Bessel functions\,\footnote{If $\nu$ is an integer, we need to use $J_{|\nu|}$ and $Y_{|\nu|}$. And $Y_{|\nu|}(x)$ is not normalizable at $x=0$.}
\begin{equation}
\tilde{a}_x=C_1\sqrt{\xi}J_{-\nu}(\omega\xi)+C_2\sqrt{\xi}J_\nu(\omega\xi)\sim C_1\xi^{1/2-\nu}+C_2\xi^{1/2+\nu}.
\label{eq:axsol}
\end{equation}
As a comparison, if the IR limit is at $\xi\to\infty$, we will have infalling wave by a Hankel function. In our case, however, the boundary condition is normalizability. See \cite{Kiritsis:2015oxa} for more examples about this type of boundary conditions. There are three cases:
\begin{itemize}
\item If $0<|\nu|<1/2$, both solutions are normalizable, and $\tilde{V}\to -\infty$. There is an ambiguity to specify a boundary condition in the IR. The conductivity is gapless.
\item If $1/2<|\nu|<1$, both solutions are normalizable, and $\tilde{V}\to\infty$. There is also an ambiguity to specify a boundary condition in the IR. The conductivity is gapped if $\tilde{V}\to\infty$ in the UV (when $n\geq 3$).
\item If $|\nu|\geq 1$, only one solution is normalizable, and $\tilde{V}\to\infty$. The boundary condition in the IR is uniquely specified by $C_1=0$. The conductivity is gapped if $\tilde{V}\to\infty$ in the UV (when $n\geq 3$). For the AdS soliton, we have $\nu=1$.
\end{itemize}

If the boundary condition in the IR is uniquely defined regardless of how to resolve the singularity, we require $|\nu|\geq 1$ \cite{Charmousis:2010zz,Kiritsis:2015oxa}.  For the AdS$_4$ ($n=2$) solution \eqref{eq:sol}, we always have $|\nu|>1$ under the null energy condition of the boundary CFT. For higher dimensions, $|\nu|\geq 1$ will give nontrivial constraints to the parameters in general.

Some remarks about the DC conductivity are as follows. The DC conductivity can be calculated by the membrane paradigm \cite{Iqbal:2008by}. At $\omega=0$, there is a radially conserved quantity satisfying $\Pi'=0$:
\begin{equation}
\Pi=\frac{f^{1-\beta}}{r^{n-2}}a_x'.
\end{equation}
Substituting here the solution \eqref{eq:axsol}, we obtain that $\Pi$ is a constant in the IR. Thus, $\Pi$ is the same constant at the AdS boundary, and $\Pi\sim a_x^{(n-1)}$. We can take the source of the perturbation $a_x^{(0)}=1$, and by \eqref{eq:sigma}, $\Pi(\omega=0)$ is the Drude weight of the conductivity. This means that the conductivity at $\omega=0$ has a delta function, despite the system is neutral.\footnote{It is less understood that the conductivity of a zero density system can have a delta function at $\omega=0$. This can only happen at zero temperature. There is an example from AdS$_5$ supergravity in \cite{DeWolfe:2012uv}, and two examples from AdS$_4$ supergravity in \cite{Kiritsis:2015oxa}.}

{\it Laplacian.} We can study other two-point correlation functions in the same manner. Consider tensor perturbations, which satisfy the Laplace equation. The Laplace equation $\nabla^2(e^{-i\omega t}\Phi)=0$ is
\begin{equation}
\Phi''+\left(\frac{f'}{f}-\frac{n}{r}\right)\Phi'+\frac{\omega^2}{f^{1+\alpha}}\Phi=0.
\end{equation}
Similarly, after a change of variables by
\begin{equation}
\frac{d\xi}{dr}=\frac{1}{f^{(1+\alpha)/2}},\qquad \tilde{\Phi}=r^{-\frac{n}{2}}f^{-\frac{\alpha-1}{4}}\Phi,
\end{equation}
we obtain a Schr\"{o}dinger equation \eqref{eq:schr}. The leading order of the Schr\"{o}dinger potential in the IR is
\begin{equation}
\tilde{V}=\frac{-1/4}{\xi^2},
\end{equation}
which is independent of either the dimension or other parameters. This corresponds to the $\nu=0$ case of above, and thus one solution will contain a log term. Therefore, the boundary condition in the IR given by the normalizability is unambiguous and well-defined.

\section{Relation to hyperscaling-violating geometries}
\label{sec:hyperscaling}
We will show that after a dimensional reduction of a special case of \eqref{eq:soln}, the IR geometry is equivalent to a hyperscaling-violating geometry
\begin{equation}
ds^2=\tilde{r}^\frac{2\theta}{d}\left(-\frac{dt^2}{\tilde{r}^{2z}}+\frac{d\tilde{r}^2+dx_1^2+\cdots+dx_d^2}{\tilde{r}^2}\right),
\label{eq:ztheta}
\end{equation}
where $z$ is the Lifshitz scaling exponent, $\theta$ is the hyperscaling violation exponent, and $d$ is the number of spatial dimensions in the AdS boundary.

We will examine a special case of \eqref{eq:soln}:
\begin{equation}
ds^2=\frac{1}{r^2}\left[-f^\alpha dt^2+\frac{dr^2}{f}+f^\beta(dx_1^2+\cdots+dx_d^2)+f^\gamma dy^2\right],
\label{eq:special}
\end{equation}
where
\begin{equation}
\alpha+d\beta+\gamma=1,\qquad \alpha^2+d\beta^2+\gamma=0.
\end{equation}
We assume $\alpha>\beta>0$, which requires $\gamma<0$. If we compactify the coordinate $y$ on $S^1$, and do a KK reduction, we obtain a spatially isotropic geometry, which has a hyperscaling-violating geometry in the IR. The reduction ansatz is written as
\begin{equation}
ds^2=e^{2\bar{\alpha}\phi}d\bar{s}_{d+2}^2+e^{2\bar{\beta}\phi}dy^2,
\end{equation}
where $\bar{\alpha}=1/\sqrt{2d(d+1)}$ and $\bar{\beta}=-d\bar{\alpha}$ \cite{Pope}. For more details about the KK reduction, see appendix~\ref{sec:KK}. The action after the dimensional reduction is
\begin{equation}
\bar{S}=\int d^{d+2}\sqrt{-\bar{g}}\left(\bar{R}+\frac{(d+1)(d+2)}{L^2}e^{2\bar{\alpha}\phi}-\frac{1}{2}(\partial\phi)^2\right).
\end{equation}

The dilaton is determined by $e^{2\bar{\beta}\phi}=f^\gamma/r^2$, and the geometry is 
\begin{equation}
d\bar{s}^2=\frac{f^{\gamma/d}}{r^{2+2/d}}\left(-f^\alpha dt^2+\frac{dr^2}{f}+f^\beta(dx_1^2+\cdots+dx_d^2)\right).
\end{equation}
The boundary is conformal to Minkowski space. After we take the IR limit $r\to 1$, and define a new variable $\bar{r}=\sqrt{1-r}$, we obtain an IR geometry
\begin{equation}
ds^2=\bar{r}^{2\gamma/d}\left[-\bar{r}^{2\alpha} dt^2+d\bar{r}^2+\bar{r}^{2\beta}(dx_1^2+\cdots+dx_d^2)\right],\label{eq:redgeo}
\end{equation}
where $\bar{r}\to 0$ is the IR limit. This is a hyperscaling-violating geometry. The coordinate $\bar{r}$ is related to the $\tilde{r}$ in \eqref{eq:ztheta} by
\begin{equation}
\bar{r}^{\gamma/d+\beta}=\tilde{r}^{\theta/d-1}.
\end{equation}
By comparing the two metrics, we obtain
\begin{equation}
z=\frac{1-\alpha}{1-\beta},\qquad \theta=\frac{\gamma+d}{1-\beta}.
\end{equation}
The following relation between $z$ and $\theta$ is always satisfied
\begin{equation}
\theta=z+d.
\end{equation}

A consequence of the relation $\theta=z+d$ is that the hyperscaling-violating geometry \eqref{eq:redgeo} cannot be regarded as an extremal limit of a finite temperature geometry. If a hyperscaling-violating geometry is continuously connected to a finite temperature geometry, i.e., the Gubser criterion is satisfied, the extremal geometry is given by
\begin{equation}
ds^2=\frac{1}{r^2}\left(-r^\frac{2d(z-1)}{d-\theta}fdt^2+r^\frac{2\theta}{d-\theta}\frac{dr^2}{f}+\sum_{i=1}^ddx_i^2\right),
\end{equation}
where
\begin{equation}
f=1-\left(\frac{r}{r_h}\right)^\frac{d(d+z-\theta)}{d-\theta}.
\end{equation}
The horizon is at $r=r_h$, and $0<r<r_h$. The extremal limit is at $r_h\to\infty$. The near-extremal solution exists only when $(d-\theta)(d+z-\theta)>0$. In our case, the Gubser criterion is marginally violated, and the near-extremal solution does not exist.

The null energy condition for the hyperscaling-violating geometry is \cite{Dong:2012se}
\begin{equation}
(d-\theta)(d(z-1)-\theta) \geq 0,\qquad
(z-1)(d+z-\theta) \geq 0.
\end{equation}
These inequalities are always satisfied in our case; the second one takes the equal sign.


In the IR limit, the compactified $S^1$ will have an infinitely large radius. We can analyze a more general dimensional reduction, for example
\begin{equation}
ds^2=\frac{1}{r^2}\left(-f^\alpha dt^2+\frac{dr^2}{f}+\sum_{i=1}^{d}f^{\beta_i}dx_i^2+\sum_{i=1}^{d'}f^{\gamma_i} dy_i^2\right),
\end{equation}
where $\alpha>0$ and $\beta_i>0$. After a KK reduction on a $d'$-dimensional torus $T^{d'}$, we will obtain a $(d+2)$-dimensional geometry. Similarly, the compactified $T^{d'}$ will have an infinitely large size in the IR limit.

\section{Discussion}
\label{sec:dis}
We have studied anisotropic solutions to Einstein's equations in AdS$_4$ and higher dimensions. They describe a holographic RG flow from an asymptotically AdS boundary to a timelike Kasner singularity, which is a less generic case of the BKL singularity. The key conclusions are as follows.
\begin{itemize}
\item The null energy condition for the boundary CFT requires $p_t\geq p_i$ for all $i$. This ensures $g_{tt}\to 0$ in the IR limit. As the IR is approached, some spatial directions are contracting ($g_{ii}\to 0$), and some spatial directions are expanding ($g_{ii}\to\infty$), but the product of all spatial $g_{ii}$ vanishes.
\item The causality constraint is always satisfied, and there is no barrier for extremal surfaces to probe the IR.
\item The boundary CFT is in a non-confining phase for the anisotropic AdS$_4$, but not for higher dimensions in general.
\item In the IR, we do not have infalling wave boundary conditions, instead, the boundary condition is imposed by normalizability. In this aspect, the solution is more similar to the AdS soliton than to the AdS black hole. For AdS$_4$, there is always only one normalizable solution in the IR, but for higher dimensions, a unique normalizable solution will give nontrivial constraints to the parameters.
\item In a spacial case, we obtain a hyperscaling-violating geometry after a KK reduction. The geometry is not connected to a finite temperature black hole.
\end{itemize}

A possible scenario to remedy the singularity is as follows. Starting from a gravity system coupled to matter fields \cite{Iizuka:2012wt}, the system admits an anisotropic solution with a regular horizon. Then as the extremal limit is taken, the matter fields vanish,\footnote{We do have some supergravity systems in which the gauge field vanishes as the extremal limit is taken, for example, the 1-charge and 2-charge black holes in the appendix~C of \cite{Kiritsis:2015oxa}.} and the geometry is described by the anisotropic solution. In a weaker case, if the matter fields are irrelevant in the IR, we can get a Kasner singularity in the IR.

Further investigations of the anisotropic solution are necessary. Are there further evidence to justify the naked singularity? What is the field theory dual to the anisotropic solution, especially for the AdS$_4$ and AdS$_5$ cases? Other questions are as follows.
\begin{itemize}
\item It will be interesting if more general analytic solutions can be obtained. For example, can we generalize the solution with an electromagnetic field?
\item The solution \eqref{eq:sol} is obtained by the Bianchi Type I Universe, but are there counterparts of other Bianchi types of the Universe?
\item The IR geometry in this work is the less generic case of the timelike BKL singularity. How to construct an RG flow from the asymptotically AdS boundary to the generic case of the timelike BKL singularity?
\item The time evolution of the Universe is mapped to a holographic RG flow. There are cosmological models to avoid the Big Bang singularity, such as bouncing models. Can we resolve the singularity in the IR analogously?
\end{itemize}

\vspace{5pt}
{\it Some remarks.} A special case of the AdS$_5$ anisotropic solution in Fefferman-Graham coordinates has appeared in \cite{Janik:2008tc} to study anisotropic plazmas at strong coupling; the solution looks complicated and does not have an explicit relation to the Kasner geometry. Both \cite{Janik:2008tc} and a later work \cite{Rebhan:2011ke} conclude that there exists infalling wave at the IR. According to our detailed analysis in section~\ref{sec:conductivity}, there is no infalling wave in the IR, but we still have normalizability as a well-defined boundary condition.

There is a large liturature about anisotropic plasmas, notably \cite{Mateos:2011ix,Mateos:2011tv}; we did not address these topics. Matter fields are involved in these geometries.
There are anisotropic solutions in other contexts, for example, \cite{Pal:2009yp,Iizuka:2012pn}.

\acknowledgments
I thank Steve Gubser, Chris Herzog, Elias Kiritsis, and Li Li for helpful discussions. This work is partially supported by the American-Israeli Bi-National Science Foundation, the Israel Science Foundation Center of Excellence and the I-Core Program of the Planning and Budgeting Committee and The Israel Science Foundation ``The Quantum Universe''.

\appendix
\section{Relation to other coordinates}
\label{sec:other}
The solution \eqref{eq:sol} has three commuting Killing vectors, which implies that it is related to a Bianchi type I Universe after a double Wick rotation and appropriate coordinate transformations. We will consider (3+1)-dimensional spacetimes, and the result can be straightforwardly generalized to arbitrary dimensions.

In the Ricci-flat case, we have the Kasner solution \cite{Kasner:1921} for a Bianchi type I Universe:
\begin{equation}
ds^2=-dt^2+\sum_{i=1}^3t^{2p_i}dx_i^2,\label{eq:kasnert0}
\end{equation}
where $p_i$ satisfy the Kasner conditions \eqref{eq:kasner}. If the geometry has a cosmological constant, the above solution can be generalized to \cite{Stephani:2003,Kasner:1925}
\begin{equation}
ds^2=-dt^2+G(t)^{2/3}\sum_{i=1}^3e^{2(p_i-1/3)U(t)}dx_i^2,\label{eq:kasnert}
\end{equation}
where
\begin{equation}
\dot{U}=1/G,\qquad \dot{G}^2+3\Lambda G^2=1=\sum_{i=1}^3p_i=\sum_{i=1}^3p_i^2.
\end{equation}
After a double Wick rotation ($\epsilon_0=-1$ and $\epsilon_{1,2}=1$), we obtain \cite{Stephani:2003,MacCallum:1997cn}
\begin{equation}
ds^2=dr^2+G(r)^{2/3}\sum_{i=0}^2\epsilon_ie^{2(p_i-1/3)U(r)}dx_i^2,\label{eq:kasnerr}
\end{equation}
where
\begin{equation}
U'=1/G,\qquad G'^2+3\Lambda G^2=1=\sum_{i=0}^2p_i=\sum_{i=0}^2p_i^2.
\end{equation}
There are other cases for $G$; please read \cite{MacCallum:1997cn} for details.

We consider a negative cosmological constant here. The cosmological constant and the AdS$_4$ radius are related by $-2\Lambda=6/L^2$. We can use the following solutions for $G(r)$ and $U(r)$:
\begin{equation}
G=\frac{1}{\sqrt{3|\Lambda|}}\sinh\left(\sqrt{3|\Lambda|}\,r\right),\qquad e^U=\sqrt{\frac{|\Lambda|}{3}}\tanh\left(\frac{1}{2}\sqrt{3|\Lambda|}\,r\right).
\end{equation}
After the coordinate transformation
\begin{equation}
1+\cosh\left(\sqrt{3|\Lambda|}\,r\right)=|\Lambda|\,\bar{r}^3,
\end{equation}
we obtain the following metric with $m=1/3$:
\begin{equation}
ds^2=-fh^{\alpha-1}dt^2+\frac{d\bar{r}^2}{f}+\bar{r}^2\left(h^\beta dx^2+h^\gamma dy^2\right),
\end{equation}
where
\begin{eqnarray}
f(\bar{r}) &=& -\frac{2m}{\bar{r}}+\frac{\bar{r}^2}{L^2},\\
h(\bar{r}) &=& \frac{f(\bar{r})}{\bar{r}^2}.
\end{eqnarray}
When $\alpha=1$, the solution is the AdS black hole, and when $\alpha=0$, the solution is the AdS soliton
\begin{equation}
ds^2=-\bar{r}^2dt^2+\frac{d\bar{r}^2}{f(\bar{r})}+f(\bar{r})dx^2+\bar{r}^2dy^2.
\end{equation}
By defining $r=1/\bar{r}$ and the freedom of rescaling the AdS radius and the horizon size, we can obtain the solution \eqref{eq:sol}.

The anisotropic solution in Poincar\'{e} coordinates has a remarkably simple form with a cosmological constant, unlike \eqref{eq:kasnert} or \eqref{eq:kasnerr}, which are much more cumbersome than the case without a cosmological constant. Moreover, the solution \eqref{eq:sol} has an explicit relation between the UV and the IR.

\section{Notes for an integral}
\label{sec:int}
The calculation of the holographic entanglement entropy and Wilson loops in section~\ref{sec:Wilson} involves the following integral:
\begin{equation}
l(r_m)=2r_m\int_0^1\frac{u^k\left(1-r_m^du^d\right)^b}{\sqrt{\left(\dfrac{1-r_m^du^d}{1-r_m^d}\right)^a-u^4}}\,du,
\label{eq:lrm}
\end{equation}
where $0\leq r_m\leq 1$. Obviously, $l(r_m=0)=0$. We want to know whether $l(r_m)$ can be divergent. At the maximal value of $r_m$, there are three cases as follows:
\begin{itemize}
\item $a=0$. The leading order of $l(r_m)$ as $r_m\to 1$ is
\begin{equation}
l(r_m)=2\int_0^1\frac{u^k(1-u^d)^b}{\sqrt{1-u^4}}\,du \,\sim\, \int_{u\approx 1}(1-u)^{b-1/2}du.
\end{equation}
It is divergent when $b\leq -1/2$, and finite when $b>-1/2$.

\item $a>0$. The leading order of $l(r_m)$ as $r_m\to 1$ is
\begin{equation}
l(r_m)=2(d(1-r_m))^{a/2}\int_0^1 u^k(1-u^d)^{b-a/2}du.
\end{equation}
It is zero at $r=r_m$, since $a>0$.

\item $a<0$. The maximal value of $r_m$ cannot reach $1$, because what is inside the square root cannot be negative. We will have $l(r_m)\to\infty$ at the maximal value of $r_m$.
\end{itemize}

\section{Kaluza-Klein reduction on $S^1$}
\label{sec:KK}
We start from the $(D+1)$-dimensional action
\begin{equation}
S=\int d^{D+1}x\sqrt{-g}\,(R-2\Lambda).
\end{equation}
We write the $(D+1)$-dimensional metric as
\begin{equation}
ds_{D+1}^2=e^{2\bar{\alpha}\phi}d\bar{s}_D^2+e^{2\bar{\beta}\phi}(dz+A)^2,\label{eq:redans}
\end{equation}
where $d\bar{s}^2$ is the $D$-dimensional metric. The $D$-dimensional action is
\begin{equation}
\bar{S}=\int d^Dx\sqrt{-\bar{g}}\left(\bar{R}-2\Lambda e^{2\bar{\alpha}\phi}-\frac{1}{2}(\partial\phi)^2-\frac{1}{4}e^{-2(D-1)\bar{\alpha}\phi}F^2\right).\label{eq:redlag}
\end{equation}
In the above expressions,
\begin{equation}
\bar{\alpha}=\frac{1}{\sqrt{2(D-1)(D-2)}},\qquad \bar{\beta}=-(D-2)\bar{\alpha}.
\end{equation}

We can also obtain an Einstein-Maxwell-dilaton (EMD) system from the dimensional reduction. After applying the following Lorentz boost to the metric \eqref{eq:special}:
\begin{equation}
t\to ct+sy,\qquad y\to st+cy,
\end{equation}
where $s\equiv\sinh\theta$ and $c\equiv\cosh\theta$, the metric becomes
\begin{equation}
ds^2=\frac{1}{r^2}\left[-\frac{f^{\alpha+\gamma}}{-f^\alpha s^2+f^\gamma c^2}dt^2+\frac{dr^2}{f}+f^\beta d\mathbf{x}^2+(-f^\alpha s^2+f^\gamma c^2)\Bigl[dy+\frac{cs(-f^\alpha+f^\gamma)}{-f^\alpha s^2+f^\gamma c^2}dt\Bigr]^2\right],
\end{equation}
where $d\mathbf{x}^2$ has $d$ dimensions. The Lagrangian after the dimensional reduction is
\begin{equation}
\bar{S}=\int d^{d+2}\sqrt{-\bar{g}}\left(\bar{R}+\frac{(d+1)(d+2)}{L^2}e^{2\bar{\alpha}\phi}-\frac{1}{2}(\partial\phi)^2-\frac{1}{4}e^{-2(d+1)\bar{\alpha}\phi}F^2\right),\label{eq:EMD}
\end{equation}
where $F=dA$. The dilaton is
\begin{equation}
e^{-2d\bar{\alpha}\phi}=\frac{-f^\alpha s^2+f^\gamma c^2}{r^2}.
\end{equation}
The metric is
\begin{equation}
d\bar{s}^2=\frac{(-f^\alpha s^2+f^\gamma c^2)^{1/d}}{r^{2+2/d}}\left[-\frac{f^{\alpha+\gamma}}{-f^\alpha s^2+f^\gamma c^2}dt^2+\frac{dr^2}{f}+f^\beta d\mathbf{x}^2\right].
\end{equation}
The gauge field is
\begin{equation}
A_t=\frac{cs(-f^\alpha+f^\gamma)}{-f^\alpha s^2+f^\gamma c^2}.
\end{equation}

We can treat \eqref{eq:EMD} as the starting point of the model, and calculate the conductivity. We turn on the perturbations $\delta A_t=e^{-i\omega t}a_x$ and $\delta g_{tx}=e^{-i\omega t}h_{tx}$ \cite{Hartnoll:2008kx}. After eliminating $h_{tx}$, the equation for $a_x$ is
\begin{equation}
a_x''+\frac{(Z\sqrt{-g}\,g^{xx}g^{rr})'}{Z\sqrt{-g}\,g^{xx}g^{rr}}a_x'+\left(\frac{\omega^2}{|g_{tt}|g^{rr}}-\frac{ZA_t'^2}{|g_{tt}|}\right)a_x=0.\label{eq:axgeneral}
\end{equation}
After a change of variables by
\begin{equation}
\frac{d\xi}{dr}=\left(\frac{-f^\alpha s^2+f^\gamma c^2}{f^{1+\alpha+\gamma}}\right)^\frac{1}{2},\qquad \tilde{a}_x=r^{-\frac{d+1}{2}}f^{-\frac{\alpha+2\beta+\gamma-1}{4}}(-f^\alpha s^2+f^\gamma c^2)^\frac{1}{2}a_x,
\end{equation}
we obtain a Schr\"{o}dinger equation \eqref{eq:schr}. The IR limit is at $\xi\to 0$. The leading order of the Schr\"{o}dinger potential is
\begin{equation}
\tilde{V}=\frac{\nu^2-1/4}{\xi^2},\qquad \nu=\frac{2\beta-\gamma}{2(1-\alpha)},
\end{equation}
which is independent of the boost rapidity. For $d=1$ ($D=3$), we always have $|\nu|>1$, and thus the boundary condition in the IR is uniquely defined.

\end{document}